\documentclass[prb,twocolumn,superscriptaddress,showpacs,letterpaper]{revtex4}
\usepackage{graphicx}
\begin{document}
\newtheorem{theorem}{Theorem}
\newtheorem{acknowledgement}[theorem]{Acknowledgement}
\newtheorem{algorithm}[theorem]{Algorithm}
\newtheorem{axiom}[theorem]{Axiom}
\newtheorem{claim}[theorem]{Claim}
\newtheorem{conclusion}[theorem]{Conclusion}
\newtheorem{condition}[theorem]{Condition}
\newtheorem{conjecture}[theorem]{Conjecture}
\newtheorem{corollary}[theorem]{Corollary}
\newtheorem{criterion}[theorem]{Criterion}
\newtheorem{definition}[theorem]{Definition}
\newtheorem{example}[theorem]{Example}
\newtheorem{exercise}[theorem]{Exercise}
\newtheorem{lemma}[theorem]{Lemma}
\newtheorem{notation}[theorem]{Notation}
\newtheorem{problem}[theorem]{Problem}
\newtheorem{proposition}[theorem]{Proposition}
\newtheorem{remark}[theorem]{Remark}
\newtheorem{solution}[theorem]{Solution}
\newtheorem{summary}[theorem]{Summary}   
\def\r{{\bf{r}}}
\def\i{{\bf{i}}}
\def\j{{\bf{j}}}
\def\m{{\bf{m}}}
\def\k{{\bf{k}}}
\def\kt{{\tilde{\k}}}
\def\mt{{\hat{t}}}
\def\mG{{\hat{G}}}
\def\mg{{\hat{g}}}
\def\mGa{{\hat{\Gamma}}}
\def\mS{{\hat{\Sigma}}}
\def\mT{{\hat{T}}}
\def\K{{\bf{K}}}
\def\P{{\bf{P}}}
\def\q{{\bf{q}}}
\def\Q{{\bf{Q}}}
\def\p{{\bf{p}}}
\def\x{{\bf{x}}}
\def\X{{\bf{X}}}
\def\Y{{\bf{Y}}}
\def\F{{\bf{F}}}
\def\G{{\bf{G}}}
\def\bG{{\bar{G}}}
\def\mbG{{\hat{\bar{G}}}}
\def\M{{\bf{M}}}
\def\V{\cal V}
\def\tchi{\tilde{\chi}}
\def\tx{\tilde{\bf{x}}}
\def\tk{\tilde{\bf{k}}}
\def\tK{\tilde{\bf{K}}}
\def\tq{\tilde{\bf{q}}}
\def\tQ{\tilde{\bf{Q}}}
\def\si{\sigma}
\def\ep{\epsilon}
\def\hep{{\hat{\epsilon}}}
\def\al{\alpha}
\def\be{\beta}
\def\ep{\epsilon}
\def\bep{\bar{\epsilon}_\K}
\def\up{\uparrow}
\def\de{\delta}
\def\De{\Delta}
\def\up{\uparrow}
\def\dwn{\downarrow}
\def\ksi{\xi}
\def\etha{\eta}
\def\product{\prod}
\def\goto{\rightarrow}
\def\switch{\leftrightarrow}
\title{Effect of Inhomogeneity on s-wave Superconductivity in the Attractive Hubbard Model} 

\author{K.~Aryanpour}
\affiliation{Department of Physics, University of California, 
Davis, California 95616} 

\author{E.~R.~Dagotto}
\affiliation{Department of Physics and Astronomy,
University of Tennessee, 
Knoxville, TN 37996, and
Condensed Matter Sciences Division, Oak Ridge 
National Laboratory, Oak Ridge,
Tennessee 37831} 

\author{M.~Mayr} 
%\affiliation{Max-Plank-Institut f\"ur Festk\"orperforschung,
% 70569 Stuttgart, Germany}
\affiliation{Department of Physics and Astronomy,
University of Tennessee, 
Knoxville, TN 37996, and
Condensed Matter Sciences Division, Oak Ridge 
National Laboratory, Oak Ridge,
Tennessee 37831} 

\author{T.~Paiva} 
\affiliation{Departamento de Fisica dos S\'olidos,
Instituto de Fisica, Universidade Federal do Rio de Janeiro, 
Cx.P. 68528, 21945-970, Rio de Janeiro, RJ, Brazil}

\author{W.~E.~Pickett}
\affiliation{Department of Physics, University of California, 
Davis, California 95616}

\author{R.~T.~Scalettar} 
\affiliation{Department of Physics, University of California, 
Davis, California 95616}

\date{\today}
\begin{abstract}
Inhomogeneous s-wave superconductivity is studied in the two-dimensional,
square lattice attractive Hubbard Hamiltonian using the 
Bogoliubov-de Gennes (BdG) mean field approximation. We find that 
at weak coupling, and for densities mainly below half-filling, an inhomogeneous interaction in which the on-site interaction $U_i$ takes on two values, $U_i=0, 2U$ results in a larger zero temperature pairing amplitude, and that the superconducting $T_c$ can also be significantly increased, relative to a uniform system with $U_i=U$ on all sites. These effects are observed for stripe, checkerboard, and even random patterns of the attractive centers, suggesting that the pattern of inhomogeneity is unimportant. Monte Carlo calculations which reintroduce some of the fluctuations neglected within the BdG approach see the same effect, both for the attractive Hubbard model and a Hamiltonian with d-wave pairing symmetry.
\end{abstract}
%\pacs{}
%
\maketitle 
\section{Motivation}
\label{sec:introduction}            
\par One of the main themes of recent studies of strongly correlated electronic
systems is the importance of spatial inhomogeneities.
These can result either from intrinsic quenched disorder in the system,
as in the metal-insulator transition in two dimensions\cite{Lee85,Belitz94},
or arise spontaneously in an otherwise translationally invariant system.
For example, holes doped into the high temperature superconductors, 
(HTS) appear not to spread out uniformly in the CuO$_{2}$ 
planes, but instead arrange themselves in the form of stripes, 
checkerboard or perhaps even more exotic 
structures.\cite{mcelroy,hanaguri,vershinin,mook,tranquada} 
Besides the cuprate 
superconductors, such spatially varying density and spin structures are 
also key features in the physics of the manganites \cite{renner,burgy} 
and cobaltites.\cite{foo,lee}
\par Considerable theoretical work on the interplay between spatial
inhomogeneity, magnetism, and superconductivity has utilized 
the repulsive
Hubbard and t-J Hamiltonians.\cite{zaanen,machida,kato,white,vojta,seibold,kivelson} For the $2D$ square lattice
these models are known to display antiferromagnetism at 
half-filling, and, although it is less certain, perhaps also 
d-wave superconductivity when doped.  There is considerable 
evidence that they also might possess inhomogeneous stripe or 
checkerboard ground states.\cite{white,vojta}
 While DMRG treatments \cite{white} provide detailed information on the real space charge, spin, and pairing orders, the precise nature of the interplay, and whether the different orders compete or cooperate, remains unclear. In addition, the enhancement of the superconducting transition temperature $T_c$ by local inhomogeneity has been demonstrated by Martin {\it et al}.\cite{martin} 
\section{Model and Methodology}
\label{sec:formalism}
\par In this paper we address the general issue of whether inhomogeneous
regions of attraction favor superconductivity relative to the homogeneous system with the same average attraction, either by increasing
the zero temperature pairing amplitude or the transition temperature.
In many of the systems for which this question is fundamental,
such as the cuprate superconductors mentioned above, the situation
is complicated by the presence of other types of order
such as antiferromagnetism, exotic spin-gap phases, 
and nontrivial d-wave symmetry of the superconducting order parameter.
Rather than using a model like the repulsive Hubbard Hamiltonian
which incorporates this full richness,
it is useful to study the problem first in a more simple context.
Here we will present a solution of the inhomogeneous 
Bogoliubov-de Gennes (BdG) equations for the attractive Hubbard 
Hamiltonian,
\begin{eqnarray}
\label{eq:attr-hub-mod}
H=&-&t\sum_{<ij>,\sigma}(c^{\dag}_{i\sigma}c_{j\sigma} +
c^{\dag}_{j\sigma}c_{i\sigma})
\nonumber \\
&-& \mu\sum_{i\sigma}c^{\dag}_{i\sigma}c_{i,\sigma}-\sum_{i}\big|U_{i}\big|
n_{i\uparrow}n_{i\downarrow}\,,
\end{eqnarray}
with $t$ the hopping amplitude, $\mu$ the chemical potential and $U_{i}$ the local attractive interaction between the fermions of opposite spins residing on the same lattice site $i$. Our focus will be on inhomogeneous patterns in the interaction $U_{i}$. The interaction in the attractive Hubbard model can be thought of as a phenomenological one, originating, for example, from integrating out a local phonon mode.\cite{micnas90} The two-dimensional uniform attractive Hubbard model is known to yield degenerate superconductivity and charge density wave (CDW) long range order at half-filling and zero temperature.\cite{robaszkiewicz,shiba,emery} However, away from half-filling, the CDW pairing symmetry is broken and superconductivity is more favorable, and the superconducting phase transition is at finite temperature.
\par Within the BdG mean field decomposition, we replace the local pairing amplitude and local density by their average values, $\Delta_{i}=\big<c_{i\uparrow}c_{i\downarrow}\big>$ and $\big<n_{i\sigma}\big>=\big<c^{\dag}_{i\sigma}c_{i\sigma}\big>$ and arrive at the quadratic effective Hamiltonian
\begin{eqnarray}
\label{eq:Heff-bdg}
{\cal{H}}_{eff}=&-&t\sum_{<ij>,\sigma}(c^{\dag}_{i\sigma}c_{j\sigma} +
c^{\dag}_{j\sigma}c_{i,\sigma}) - \sum_{i\sigma}{\tilde\mu_i}
c^{\dag}_{i\sigma}c_{i\sigma}
\nonumber \\
&-& \sum_{i} 
\big|U_{i}\big|\big[\Delta_{i}c^{\dag}_{i\uparrow}c^{\dag}_
{i\downarrow}+\Delta^{*}_{i}c_{i\downarrow}c_{i\uparrow}\big]\,,
\end{eqnarray}
where ${\tilde\mu_i}=\mu+\big|U_{i}\big| \langle n_{i} \rangle/2$ 
includes a site-dependent 
Hartree shift with $\langle n_{i} \rangle =\sum_{\sigma} 
\langle n_{i\sigma} \rangle$. 
All energies will be referenced to $t=1$. 
\par We adopt the criterion of comparing the tendency for superconductivity in the homogeneous system with the same attraction $-U$ on all lattice
sites, with cases when sites with attraction are mixed with
sites where the attraction is absent, i.e., $U_{i}=0$. \cite{martin,litak} Specifically, we have 
studied systems in which sites with attractive interaction are randomly distributed \cite{litak} or arranged in checkerboard and stripe patterns. In all three inhomogeneous patterns, exactly half of the lattice sites carry interaction, and the interacting sites 
carry twice the value of $U$ as in the case of the uniform pattern.
\begin{figure}
\includegraphics[width=3.5in]{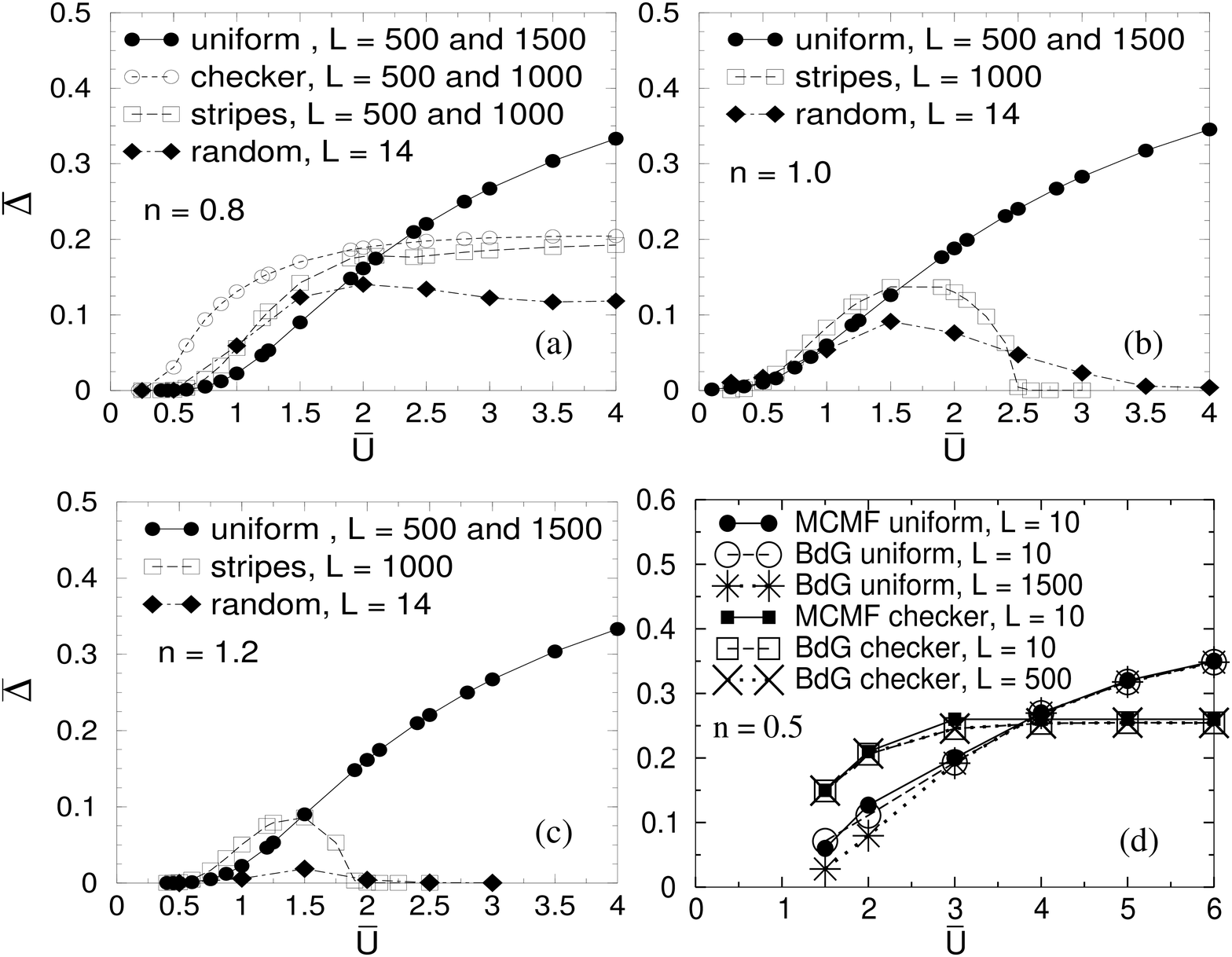}
\caption[a]{Variation of the mean zero temperature
pairing amplitude $\bar \Delta$, versus
mean interaction $\bar U$, for (a) below half-filling ($n=0.8$), (b) 
half-filling ($n=1.0$), and (c) above half-filling ($n=1.2$). 
There is no pairing at $n \geq 1$ in the checkerboard case.
For the stripe, checkerboard, and uniform patterns, data are fully
converged on the momentum space grid.  That is,
data for the indicated lattice sizes lie on top of each other.
For the random pattern it is not possible to study large lattices, and
a single size, $L=14$ is shown. Panel (d) compares the MCMF results ($T\approx 0$) at $L=10$ with their BdG counterparts at $L=10$, $500$ and $1500$ for the uniform and 
checkerboard patterns.}
\label{deltavg.vs.Uavg}
\end{figure}
\begin{figure}
\includegraphics[width=2.2in]{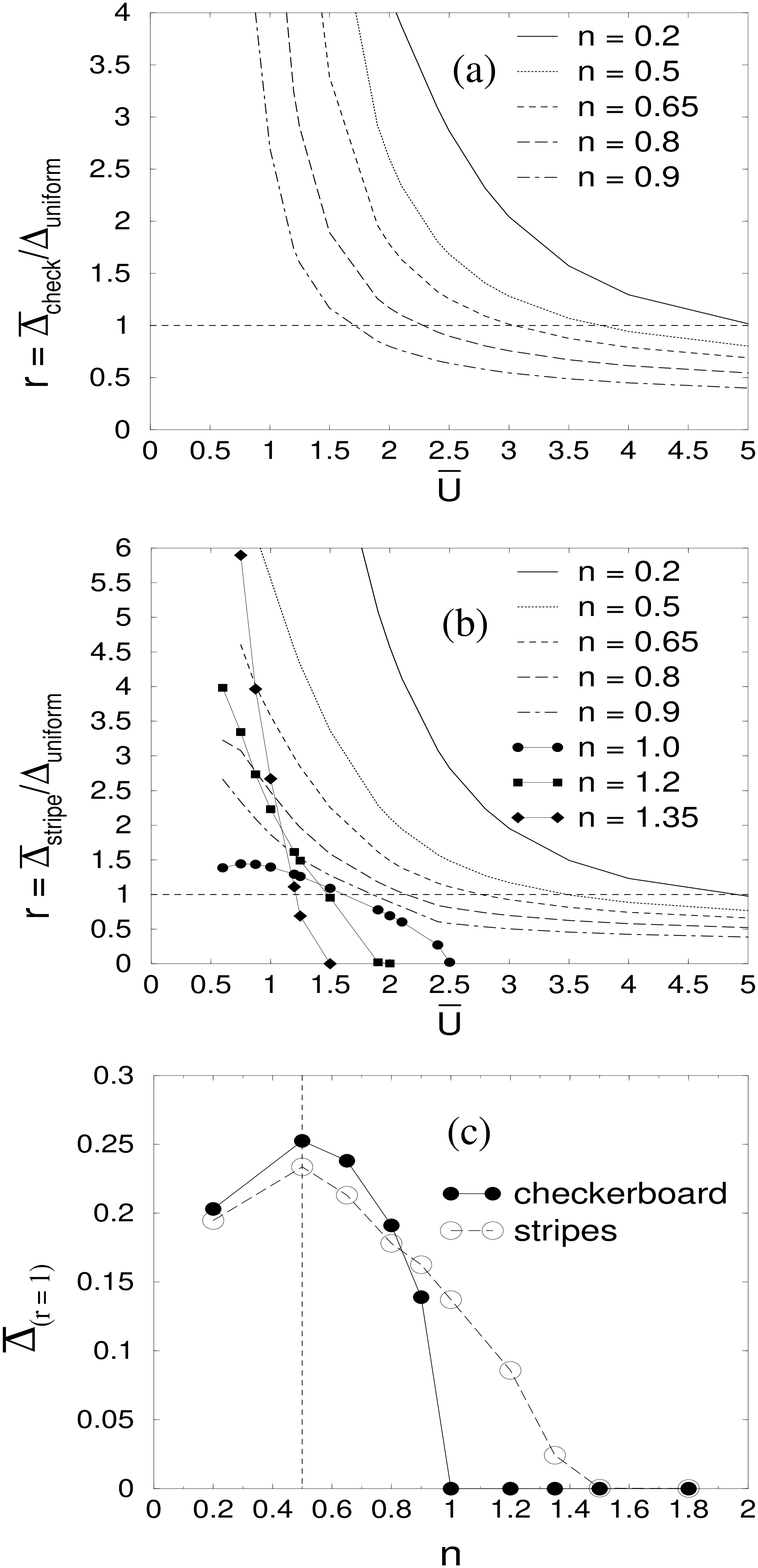}
\caption[a]{Panel (a): the ratio 
$r={\bar\Delta}_{\rm check}/{\Delta}_{\rm uniform}$ as a function of 
$\bar U$ and different dopings of electrons at $T=0$. Panel (b): 
the same results as for panel (a) for the ratio 
$r={\bar\Delta}_{\rm stripe}/{\Delta}_{\rm uniform}$. Panel (c): Value of the pairing amplitude at which $r=1$ versus the electron doping. At weak coupling,
the zero temperature pairing amplitude can be a factor of three to
four larger than the uniform system. Lattice sizes of $L=1500$ for the uniform and $L=500$ and $1000$ for inhomogeneous patterns were utilized.}
\label{enhancement}
\end{figure}
\par This conventional mean-field approach does not capture the
Kosterlitz-Thouless nature of the phase transition in two dimensions.
Nevertheless, this weakness can be repaired \cite{mayr} upon 
regarding the local pairing amplitudes as complex 
variables and performing a finite temperature Monte
Carlo integration over the associated amplitude and phase degrees of freedom. Unlike BCS, this Monte Carlo mean field (MCMF) approach, allows identification of the weak and strong coupling regimes via the phase correlation function. We will use this Monte Carlo technique as an independent confirmation of our results.
\section{Solutions and Phase Diagram}
\label{sec:results}
\par For the doping of $n=0.8$ at $T=0$, as depicted in Fig. \ref{deltavg.vs.Uavg}(a), an inhomogeneous system with bimodal $U_i=0, 2U$ has a larger
zero temperature gap than a uniform system with $U_i=U$ below the maximum value of $\bar U_{\rm (r=1)}\approx2.0$ with $r=\bar\Delta_{\rm inhomog}/\Delta_{\rm uniform}$. At half-filling in Fig. \ref{deltavg.vs.Uavg}(b),this effect for the stripe pattern terminates at a slightly smaller value of $\bar U_{\rm (r=1)}\approx1.5$ and at the same time, the magnitude of $\bar \Delta_{\rm (r=1)}$ has also diminished. There is an upper critical $\bar U_{c}$ above which superconductivity is obliterated by inhomogeneity. For the checkerboard pattern at half-filling in particular, $\bar \Delta$ becomes zero as one enters the insulating CDW phase of static pairs. For the doping of $n=1.2$ (above half-filling) in Fig. \ref{deltavg.vs.Uavg}(c) for random and stripe patterns and $\bar \Delta$ for the checkerboard pattern still remains infinitesimal over almost the entire range of $\bar U$. The MCMF data in Fig. \ref{deltavg.vs.Uavg}(d), provide confirmation of the BdG results. While the MCMF method remains limited to relatively small lattice sizes, BdG allows us to invoke lattice sizes as large as $L=1500$ to reduce the finite size effects at small $\bar U$ values. As observed in Fig. \ref{deltavg.vs.Uavg}(d), the finite size effects are more significant for the uniform pattern when $\bar U \lesssim 3.0$ and less severe for the checkerboard. The good agreement between these two approaches helps justify both the application and results of the BdG technique. 
\par Fig. \ref{enhancement} illustrates further the size of the effect due to inhomogeneity, using data over a broad range of densities. For the checkerboard pattern in Fig. \ref{enhancement}(a), as the doping changes from $n=0.2$ to $n=0.9$, the intersection point of the ratio $r={\bar\Delta}_{\rm check}/
{\Delta}_{\rm uniform}$ and unity line is shifted towards smaller $\bar U_{\rm (r=1)}$. Precisely at half-filling, $\bar \Delta$ for the checkerboard pattern 
vanishes. For stripes, however, as shown in Fig. \ref{enhancement}(b), the enhancement shift towards smaller $\bar U_{\rm (r=1)}$ continues through half-filling to almost $n=1.35$. Fig. \ref{enhancement}(c) shows the value of the pairing amplitude $\bar \Delta_{\rm (r=1)}$ as a function of the electron doping. Fillings around quarter-filling ($n=0.5$) have the largest ratio in pair amplitude of the bimodal $U_i=0, 2U$ to uniform $U_i=U$ interaction distributions.
\begin{figure}
\includegraphics[width=3.4in]{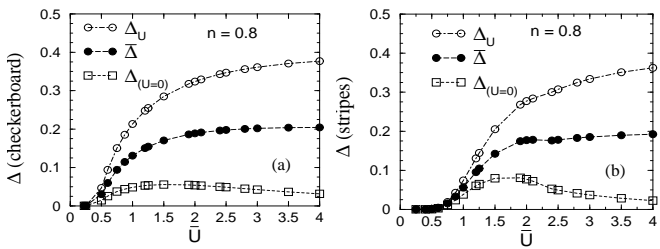}
\caption[a]{Variation of the pairing amplitude versus the mean interaction 
$\bar U$ at $T=0$. Panel (a) shows $\bar \Delta$ as the average of 
$\Delta_{\rm U}$ and $\Delta_{\rm (U=0)}$ for the checkerboard pattern 
at $n=0.8$ and the same for stripes in panel (b). 
The enhancement of  pairing by inhomogeneity is seen to arise from
a proximity effect whereby significant pair amplitude is induced on the $U=0$ sites by the $U \neq 0$ sites. Lattice sizes of $L=500$ and $1000$ were utilized.} 
\label{delta.Nf0.8.temp0.0}
\end{figure}
\par The lack of superconductivity in the 
checkerboard pattern at half-filling is associated with the formation 
of a competing, insulating CDW phase. 
This can be seen directly in the density of states as a gap develops at the 
Fermi energy despite the fact that $\Delta_{sc}=0$. The occupation of sites in real space becomes increasingly
disparate as $U$ increases, with the sites with $U \neq 0$ becoming fully
packed with $n_{\rm U}\approx2.0$, while the non-interacting sites become empty, $n_{\rm (U=0)}\approx0.0$.
For stripes at half-filling also, at large 
enough values of $\bar U$ superconductivity is obliterated. 
Above half-filling, 
however, when the order parameter vanishes for both the checkerboard and 
stripes, the density of states remains finite at the Fermi energy,
indicating a metallic phase.   
\par The proximity effect for the non-interacting sites neighbored by the interacting sites plays a major role in the magnitude of the pair amplitude in the inhomogeneous lattice. In Fig. \ref{deltavg.vs.Uavg}(a), ($n=0.8$) the value of $\bar U$ at which the checkerboard and stripe patterns have the same pair amplitude as the uniform system is $\bar U_{\rm (r=1)}\approx2.0$. In Fig. \ref{delta.Nf0.8.temp0.0}(a) and (b), $\bar \Delta$ for both these two patterns has been plotted as the average of the pairing amplitudes on interacting and non-interacting sites. Due to the proximity effect, even in the absence of interaction on a lattice site, there exists a finite value of pairing amplitude through the tunneling effect from its neighboring interacting sites. While $\Delta_{\rm U}$ for both the checkerboard and stripe patterns consistently increases as a function of $\bar U$, $\Delta_{\rm (U=0)}$  increases up to $\bar U_{\rm (r=1)}\approx2.0$ where according to Fig. \ref{deltavg.vs.Uavg}(a), the homogeneous and inhomogeneous lattices have the same pairing amplitude. $\Delta_{\rm (U=0)}$  then falls off for larger $\bar U$ values. Hence, the region of growth of $\Delta_{\rm (U=0)}$ with $\bar U$ coincides with pairing amplitude of the inhomogeneous system being large.   
\begin{figure}  
\includegraphics[width=3.4in]{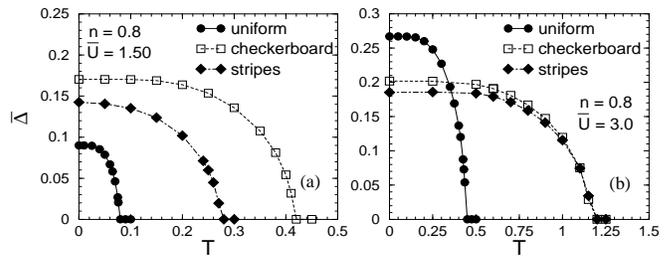}
\caption[a]{Variation of the pairing amplitude versus temperature $T$ for 
$n=0.8$. Panel (a): $\bar \Delta$ as a function of $T$ for $\bar U=1.5$ where 
checkerboard and stripe patterns with $U_i=0, 2U$ have a larger $T=0$ pair amplitude than $U_i=0$ (Fig. \ref{deltavg.vs.Uavg}(a)). Panel (b): the same results for $\bar U=3.0$. Lattice sizes of $L=1500$ for the uniform and $L=500$ and $1000$ for inhomogeneous patterns were utilized.}
\label{delta.vs.T.Nf0.8}    
\end{figure}
\par The superconducting transition temperature of a lattice with a bimodal $U_i$ can also be larger than a uniform interaction distribution. According to Fig. \ref{deltavg.vs.Uavg}(a), at $\bar U=1.5$, $\bar \Delta(T=0)$ is enhanced due to inhomogeneity but not at $\bar U=3.0$. Fig. \ref{delta.vs.T.Nf0.8} illustrates the collapse of $\bar \Delta$ as the temperature is increased for the uniform, checkerboard and stripe patterns. In Fig. \ref{delta.vs.T.Nf0.8}(a) corresponding to $\bar U=1.5$, inhomogeneity has a similar effect on the gap through all values of finite $T$ up to $T_c$ which is significantly larger for the inhomogeneous patterns. In Fig. \ref{delta.vs.T.Nf0.8}(b) corresponding to $\bar U=3.0$, at $T=0$, $\bar \Delta$ for the uniform pattern is significantly larger than its inhomogeneous counterparts. However, because the gap of the inhomogeneous system is nonzero to higher temperatures than the uniform system, for values of $0.5\lesssim T\lesssim 1.25$, there exists an enhancement region for $\bar \Delta$ for inhomogeneous patterns. For the uniform pattern, we find the $T_c$ in very good agreement with the BCS value, $k_{\rm b}T_c\approx(\Delta(0)U)/1.76$, as expected for our mean field treatment. Note that the effect of inhomogeneity on $T_c$ would appear to be even more dramatic if we compared the inhomogeneous curves against their BCS counterparts with the same $\bar \Delta(T=0)$. A similar increase in $T_c$ upon introducing a checkerboard pattern is found in the MCMF calculations as well, arising from the loss of long-range phase coherence. This is particularly significant because the MCMF incorporates the subtle nature of the superconducting transition in $2D$ discussed earlier. We have also independently confirmed that our conclusions and arguments equally apply for a model with nearest-neighbor attraction, leading to a $d$-wave SC close to half-filling, which reflects the cuprates' phenomenology more truthfully.\cite{mayr}
\section{Summary and Discussion}
\label{sec:summary}
\par In summary, we have shown that for the attractive Hubbard model on a square lattice, there is a significant range of doping and interaction strength over which the superconducting order parameter is larger for $U_i=0, 2U$ than for $U_i=U$ uniformly. It is worth emphasizing that in most situations, inhomogeneities reduce values of order parameters and critical temperatures, even when comparisons are made, as they are in this paper, to homogeneous systems with the same average value of all parameters.  This is true, for example, of classical site diluted Ising models, where the ferromagnetic $J$ is increased to compensate for absent sites, and quantum models like the boson Hubbard model where random chemical potentials monotonically decrease and ultimately destroy superfluidity.\cite{fisher89,scalettar91} An exception is the increase of $T_{\rm Neel}$ by randomness reported in DMFT studies of the repulsive model.\cite{ulmke95}
\par The increase in the SC gap has been verified for the checkerboard, stripe and random patterns and thus is insensitive to the pattern of disorder. The growth is due to the proximity effect, i.e., the tunneling effect of the Cooper pairs from the interacting sites leading to finite order parameter values even on non-interacting sites. This conclusion is supported by the effect occurring at weak coupling, where the coherence length is large, rather than in the strong coupling regime of preformed pairs. Agreement between the BdG results and the MCMF calculations justifies the application and conclusions of the BdG approach within the small $\bar U$ regime. Finally, nonuniform interaction strength also leads to the strong signal in the phase transition temperature $T_c$. Counterintuitively, this increase in $T_c$ continues even for values of $\bar U$ for which the the order parameter is larger with uniform $U_i$ than a bimodal choice. However, in this weak coupling parameter regime, $T_c$ is a supralinearly increasing function of $U$, so it may be that in the inhomogeneous system, the sites with larger $U$ produce a nonlinear enhancement relative to $T_{c}$ of the homogeneous system and, through the proximity effect, drag the $U_{i}=0$ sites along with them.
\par While the attractive Hubbard Hamiltonian obviously does not incorporate many of the features of high $T_c$ superconductors (notably the symmetry of the pairing), the model has been shown to provide useful insight into some of their phenomenology, for example the spin-gap.\cite{randeria94} It is therefore tempting to speculate that our results concerning inhomogeneity may have similar connections. Specifically, recent ARPES data \cite{yoshida03} suggests that the underdoped phase of LSCO (La$_{2-x}$Sr$_{x}$CuO$_{4}$) consists of SC clusters, embedded in the AF host. In such a system, inhomogeneous gaps appear naturally and our results here indicate that the superconducting transition is in fact determined by the largest gap values rather than the much smaller gaps found at phase boundaries, as one would naively think. This renders the SC phase more stable than it would otherwise be, and also simplifies the description of these systems.
\section{Acknowledgments}
\label{sec:acknowledgments}
\par We acknowledge useful conversations with G.~Ortiz, A.~Moreo and G.~Alvarez. This research was supported by NSF-DMR-0312261, NSF-DMR-0421810, NSF-DMR-0443144, DOE DE-FG03-03NA00071, CNPq-Brazil, FAPERJ-Brazil and FUJB-Brazil. 

\begin{thebibliography}{18}
%
\bibitem{Lee85} P.~A.~Lee and T.~V.~Ramakrishnan,
Rev. Mod. Phys. {\bf 57}, 287 (1985).
%
\bibitem{Belitz94} D.~Belitz and T.~R.~Kirkpatrick,
Rev. Mod. Phys. {\bf 66}, 261 (1994).
%
\bibitem{mcelroy} K.~McElroy, D.-H.~Lee, J.~E.~Hoffman, K.~M.~Lang, J.~Lee, E.~W.~Hudson, H.~Eisaki, S.~Uchida, and J.~C.~Davis, Phys.\ Rev.\ Lett. {\bf 94}, 197005 (2005).
%
\bibitem{hanaguri} T.~Hanaguri, C.~Lupien, Y.~Kohsaka, D.-H.~Lee, M.~Azuma, M.~Takano, H.~Takagi, and J.~C.~Davis, Nature {\bf 430}, 1001 (2004).
%
\bibitem{vershinin} M.~Vershinin, S.~Misra, S.~Ono, Y.~Abe, Y.~Ando, and A.~Yazdani, Science {\bf 303}, 1995, (2004).
%
\bibitem{mook} H.~A.~Mook, P.~Dai, and F.~Dogan, Phys.\ Rev.\ Lett. {\bf 88 }, 097004 (2002).
%
\bibitem{tranquada} J.~M.~Tranquada, J.~D.~Axe, N.~Ichikawa, A.~R.~Moodenbaugh, Y.~Nakamura, and S.~Uchida, Phys.\ Rev.\ Lett. {\bf 78}, 338 (1997).
%
\bibitem{renner} Ch.~Renner, G.~Aeppli, B.-G.~Kim, Y.-A.~Soh, and S.-W.~Cheong, Nature {\bf 416}, 518 (2002).
%
\bibitem{burgy} J.~Burgy, A.~Moreo, and E.~Dagotto, Phys.\ Rev. \ Lett. {\bf 92}, 097202 (2004).
%
\bibitem{foo} M.~L.~Foo, Y.~Wang, S.~Watauchi, H.~W.~Zandbergen, T.~He, R.~J.~Cava, and N.~P.~Ong, Phys.\ Rev.\ Lett. {\bf 92}, 247001 
(2004).
%
\bibitem{lee} K.-W.~Lee, J.~Kunes, P.~Novak, and W.~E.~Pickett, Phys.\ Rev.\ Lett. {\bf 94}, 026403 (2005).
%
\bibitem{zaanen} J.~Zaanen and O.~Gunnarsson, Phys. Rev. B {\bf 40}, R7391 (1989).
%
\bibitem{machida} K.~Machida, Physica C {\bf 158}, 192 (1989).
%
\bibitem{kato} M.~Kato, K.~Machida, H.~Nakanishi and M.~Fujita, J. Phys. Soc. Jpn., {\bf 59}, 1047 (1990).
%
\bibitem{white} S.~R.~White and D.~J.~Scalapino, Phys.\ Rev.\ B {\bf 70}, 220506(R) (2004) .
%
\bibitem{vojta} M.~Vojta, Phys. Rev. B {\bf 66}, 104505 (2002).
%
\bibitem{seibold} G.~Seibold, C.~Castellani, C.~Di~Castro and M.~Grilli,
Phys. Rev. B {\bf 58}, 13506 (1998).
%
\bibitem{kivelson} S.~A.~Kivelson and E.~Fradkin, cond-mat/0507459.
%
\bibitem{martin} I.~Martin, D.~Podolsky and S.~A.~Kivelson, Phys. Rev. B {\bf 72}, 060502(R) (2005).
%
\bibitem{micnas90} R.~Micnas, J.~Ranninger and S.~Robaskiewicz, Rev. Mod. Phys. {\bf 62}, 113 (1990) and references therein.
%
\bibitem{robaszkiewicz} S.~Robaszkiewicz, R.~Micnas, and K.~A.~Chao, Phys. Rev. B {\bf 23}, 1447 (1981). 
%
\bibitem{shiba} H.~Shiba, Prog. Theor. Phys. {\bf B48}, 2171 (1972).
%
\bibitem{emery} V.~J.~Emery, Phys. Rev. {\bf B14}, 2989 (1972).
%
\bibitem{litak} G.~Litak and B.~L.~Gy\"orffy, Phys. Rev. B {\bf 62}, 6629
 (2000).
%
\bibitem{mayr} M.~Mayr, G.~Alvarez, C.~\c Sen, and E.~Dagotto,  Phys.\ Rev.\ Lett. {\bf 94 }, 217001 (2005).
%
\bibitem{fisher89}
M.~P.~A.~Fisher, P.~B.~Weichman, G.~Grinstein, and D.~S.~Fisher, Phys. Rev. B40, 546 (1989).
%
\bibitem{scalettar91}
R.~T.~Scalettar, G.~G.~Batrouni, and G.~T.~Zimanyi, Phys. Rev. Lett. 66, 3144 (1991).
%
\bibitem{ulmke95}
M.~Ulmke, V.~Janis, D.~Vollhardt, Phys. Rev. B51, 10411 (1995).
%
\bibitem{randeria94}
M.~Randeria in {\it Bose Einstein Condensation}, A.~Giffin, D.~Snoke,
and S.~Stringari (eds), Cambridge University Press (1994), and references cited therein.
%
\bibitem{yoshida03}
T.~Yoshida, X.~J.~Zhou, T.~Sasagawa, W.~L.~Yang, P.~V.~Bogdanov, A.~Lanzara, Z.~ Hussain, T.~Mizokawa, A.~Fujimori, H.~Eisaki, Z.-X.~Shen, T.~Kakeshita, and S.~Uchida, Phys. Rev. Lett. 91, 027001 (2003).
%
\end{thebibliography}
\end{document}